\documentclass[12pt]{article}
\usepackage{graphics}
\usepackage{amssymb,amsmath}

\newcommand{\C}{\mathbb{C}}
\newcommand{\R}{\mathbb{R}}
\newcommand{\Z}{\mathbb{Z}}
\newcommand{\D}{\mathrm{d}}
\newcommand{\e}{\mathrm{e}}
\newcommand{\CC}{\mathcal{C}}

\newcommand{\OO}{\mathcal{O}}

\newcommand{\RR}{\mathcal{R}}

\newcommand{\QED}{\mbox{\rule[-.5pt]{4pt}{7.5pt}}}
\newtheorem{claim}{Claim}[section]
\newtheorem{theorem}[claim]{Theorem}

\newtheorem{proposition}[claim]{Proposition}

\newtheorem{example}[claim]{Example}

\begin{document}

\title{Inequalities for means of chords,
with application to isoperimetric problems}
\author{Pavel Exner$^{a,b}$, Evans M. Harrell$^c$, Michael Loss$^c$}
\date{}
\maketitle
\begin{flushleft}

{\small \em a) Department of Theoretical Physics, Nuclear Physics
Institute, Academy of \phantom{a) }Sciences, 25068 \v{R}e\v{z}
near Prague, Czechia; {\rm exner@ujf.cas.cz} \\ b) Doppler
Institute, Czech Technical University, B\v{r}ehov{\'a} 7, 11519
Prague \\ c) School of Mathematics, Georgia Institute of
Technology, Atlanta, Ga 30332, \phantom{a) }U.S.A.; {\rm
harrell@math.gatech.edu, loss@math.gatech.edu} }

\end{flushleft}

\vspace{8mm}

\begin{quote}
\noindent {\small We consider a pair of isoperimetric problems
arising in physics.
The first concerns a Schr\"odinger operator in $L^2(\R^2)$ with
an attractive interaction supported on a closed curve $\Gamma$,
formally given by $-\Delta-\alpha\delta(x-\Gamma)$; we ask which
curve of a given length maximizes the ground state energy. In the second
problem we have a loop-shaped thread $\Gamma$ in $\R^3$,
homogeneously charged but not conducting, and we ask about the
(renormalized) potential-energy minimizer. Both problems reduce to
purely geometric questions about
inequalities for mean values of chords of $\Gamma$.
We prove an isoperimetric theorem 
for $p$-means of chords of curves when $p \leq 2$, which
implies in particular
that the global extrema for the physical problems
are always attained when $\Gamma$ is a
circle.  The article finishes with a discussion of the
$p$--means of chords when $p > 2$.}
\end{quote}


\section{Introduction}

Isoperimetric problems are a trademark topic in mathematical
physics. In this note we consider two of them, which at first glance
are completely unrelated, and show that their solution can be
reduced to the same geometric question. The first problem
concerns a class of operators in $L^2(\R^2)$ which are given
formally by the expression
 \begin{equation} \label{formal}
 H_{\alpha,\Gamma}= -\Delta-\alpha\delta(x-\Gamma)\,,
 \end{equation}
with an attractive interaction, $\alpha>0$, where $\Gamma$ is a
smooth loop in the plane of a fixed length $L>0$; for a proper
definition of $H_{\alpha,\Gamma}$, either through the natural
quadratic form or in terms of boundary conditions on $\Gamma$ see
\cite{BEKS, EI}.

We are interested in the shape of $\Gamma$ that maximizes the
ground state energy.  There are several reasons for interest
in this question.  On the physical side, operators of the type
(\ref{formal}) have been studied as models of \emph{leaky
quantum wires}, with the aim of 
more realistically taking quantum tunneling into account.
For references 
see \cite{BT, EI}; a bibliography can be found
in \cite{AGHH2}.  Such operators exhibit
interesting connections between spectral properties and the shape of
the curve supporting the interaction, in particular, that the
curvature of $\Gamma$ generates an effective attractive
interaction.

Another inspiration is the isoperimetric problem for the
Dirichlet Laplacian on a closed loop-shaped tube, where it can be proved
that the ground state is uniquely maximized by a circular annulus
\cite{EHL}.  By a natural analogy one expects a similar result
for leaky wires, with the circle again being the maximizer. This
question was posed and partially solved in \cite{Ex2} where it was
shown that a circle maximizes the principal eigenvalue locally.
See also \cite{Ex1} for a ``discrete'' analogue of this result.
Here we present a full solution, 
demonstrating that a circular $\Gamma$ is the unique global maximizer
for the ground state of the operator $H_{\alpha,\Gamma}$.

A second motivation is to be found in classical electrostatics rather than
quantum mechanics. Suppose that a nonconducting thread 
with no rigidity
in the form of
a loop has a homogeneous charge density.  We ask which shape 
the thread will take in the
absence of other, nonelectrostatic, forces.
Roughly speaking, we seek the $\Gamma$ that minimizes
the potential energy of the electric field. The tricky part of the
question is that the potential energy is infinite.
Since the
divergence is due to short--distance effects
independent of the shape of the thread, however, the energy can be renormalized,
yielding a well-posed problem.

We shall reduce both questions to a geometric
property of chords of $\Gamma$.  A class of
inequalities for $L^p$ norms of the chords in any dimension $d\ge
2$ is proved, particular cases of which will help solve the two
problems stated above.


\setcounter{equation}{0}
\section{Mean-chord inequalities} \label{geom}

As mentioned above it is convenient to formulate the result in an
arbitrary dimension $d\ge 2$.  First we have to specify regularity
requirements on the curve. Let $\Gamma$ be a loop in $\R^d$
parametrized by its arc length, in other words, a piecewise
differentiable function $\Gamma:\: [0,L]\to \R^d$ such that
$\Gamma(0)=\Gamma(L)$ and $|\dot \Gamma(s)|=1$ for any
$s\in[0,L]$. Let us consider all the arcs of $\Gamma$ having
arc length $u\in (0,\frac12 L]$.  We shall be concerned with the
inequalities
 \begin{eqnarray}
 C_L^p(u):&\quad\;\; \int_0^L |\Gamma(s\!+\!u) -\Gamma(s)|^p\,
 \D s \,\le\, \frac{L^{1+p}}{\pi^p} \sin^p \frac{\pi u}{L}\,,&
 \; p>0\,, \label{C+p} \\
 C_L^{-p}(u):& \int_0^L |\Gamma(s\!+\!u) -\Gamma(s)|^{-p}\,
 \D s \,\ge\, \frac{\pi^p L^{1-p}}{\sin^p \frac{\pi u}{L}}\,,&
 \; p>0\,. \label{C-p}
 \end{eqnarray}
The expressions on the right sides correspond to the maximally
symmetric case, i.e., the planar circle.  It is clear that the
inequalities are invariant under scaling, so without loss of
generality we may fix the length, e.g., to $L=2\pi$. In the case
$p=0$ the inequalities (\ref{C+p}) and (\ref{C-p}) turn into
trivial identities.

A natural conjecture, to be proved in this section, is that they
hold for any $d\ge 2$ and $p \le 2$. In fact, it is sufficient to
consider only the case $p=2$ as the following simple result shows.

\begin{proposition} \label{auxil}
$C_L^p(u)$ implies $C_L^{p'}(u)$ if $p>p'>0$. Similarly,
$C_L^p(u)$ implies $C_L^{-p}(u)$ for any $p>0$.
\end{proposition}
{\sl Proof:} The first claim is due to the convexity of $x\mapsto
x^\alpha$ in $(0,\infty)$ for $\alpha>1$:  Supposing 
(\ref{C+p}), 
 \begin{eqnarray*}
 \frac{L^{1+p}}{\pi^p} \sin^p \frac{\pi u}{L} &\!\ge\!&
 \int_0^L \left(|\Gamma(s\!+\!u) -\Gamma(s)|^{p'} \right)^{p/p'}\,
 \D s \\ &\!\ge\!&
 L \left( \frac1L \int_0^L |\Gamma(s\!+\!u) -\Gamma(s)|^{p'}\,
 \D s \right)^{p/p'}\,.
 \end{eqnarray*}
It is then sufficient to take both sides to the power $p'/p$.
Furthermore, the Schwarz inequality implies
 $$ 
 \int_0^L |\Gamma(s\!+\!u) -\Gamma(s)|^{-p}\, \D s \ge
 \frac{L^2}{\int_0^L |\Gamma(s\!+\!u) -\Gamma(s)|^p\, \D s} \ge
 \frac{L^2\pi^p}{L^{1+p}\sin^p \frac{\pi u}{L}}\,,
 $$ 
which completes the proof of (\ref{C-p}). \hfill\QED \vspace{.5em}

The claim for $p=2$ can be proved by means of a simple Fourier
analysis.

\begin{theorem} \label{p=2}
Suppose that $\Gamma$ belongs to the described class.  Then $C_L^2(u)$ is
valid for any $u\in(0,\frac12 L]$, and the inequality is 
strict
unless $\Gamma$ is a planar circle.
\end{theorem}
{\sl Proof:}  Without loss of generality we
put $L=2\pi$ and write
 \begin{equation} \label{expans}
 \Gamma(s) = \sum_{0\ne n\in\Z} c_n\, \e^{ins}
 \end{equation}
with $c_n\in\C^d$; since $\Gamma(s)\in\R^d$ the coefficients have
to satisfy the condition
 $$ 
 c_{-n} = \bar{c}_n\,.
 $$ 
The absence of $c_0$ can be always achieved by a choice of the
origin of the coordinate system. In view of the Weierstrass
theorem and continuity of the functional in question, we may also
suppose that $\Gamma$ is of the class $C^2$, in which case
\cite[Sec.~VIII.1.2]{KF} its derivative is a sum of the uniformly
convergent Fourier series
 \begin{equation} \label{der-expans}
 \dot\Gamma(s) = i \sum_{0\ne n\in\Z} nc_n\, \e^{ins}\,.
 \end{equation}
By assumption, $|\dot\Gamma(s)|=1$, and hence from
the relation
 $$ 
 2\pi = \int_0^{2\pi} |\dot\Gamma(s)|^2\, \D s = \int_0^{2\pi}
 \sum_{0\ne m\in\Z}\: \sum_{0\ne n\in\Z} nm\, c^*_m\cdot c_n\,
 \e^{i(n-m)s} \,\D s\,,
 $$ 
where $c^*_m$ denotes the row vector $(\bar c_{m,1},\dots, \bar
c_{m,d})$ and dot marks the inner product in $\C^d$, we infer that
 \begin{equation} \label{norm}
 \sum_{0\ne n\in\Z} n^2 |c_n|^2 = 1\,.
 \end{equation}
In a similar way we can rewrite the right member of
$C_{2\pi}^2(u)$ using the Parseval relation as
 $$ 
 \int_0^{2\pi} \left| \sum_{0\ne n\in\Z} c_n\, (\e^{inu}-1)\,
 \e^{ins}\right|^2 \,\D s = 8\pi \sum_{0\ne n\in\Z} |c_n|^2
 \left(\sin\frac{nu}{2}\right)^2\,,
 $$ 
and thus the sought inequality is equivalent to
 \begin{equation} \label{ineq}
 \sum_{0\ne n\in\Z} n^2 |c_n|^2
 \left(\frac{\sin\frac{nu}{2}}{n \sin\frac{u}{2}}\right)^2 \le 1.
 \end{equation}
It is therefore sufficient to prove that
 \begin{equation} \label{sin-ineq}
 \left| \sin nx\right| \le n\,\sin x
 \end{equation}
for all positive integers $n$ and all $x\in(0,\frac12\pi]$. 
We proceed by induction.  The
claim is obviously valid for $n=1$.  Suppose now that
(\ref{sin-ineq}) holds for a given $n$. We have
 $$ 
 (n+1)\sin x \mp \sin (n+1)x = n\sin x \mp \sin nx \cos x +
 \sin x (1\mp \cos nx),
 $$ 
where the sum of the first two terms at the right side is
non-negative by assumption, and the same is clearly true for the
last one. 
In combination with
the continuity argument this proves $C^2_L(u)$ for the
considered class of functions.

The induction argument shows that if (\ref{sin-ineq}) is
strict
for $n$ it is 
likewise strict
for $n+1$.  Since the inequality is
strict
for $n=2$, equality can occur only for $n=1$.  This in
turn means that the inequality (\ref{ineq}) is 
strict
unless
$c_n=0$ for $|n|\ge 2$, and, consequently, $C_{2\pi}^2(u)$ is
saturated only if the projection of $\Gamma$ on the $j$th axis
equals
 $$ 
 \Gamma_j(s) = 2|c_{1,j}| \cos(s+ \arg c_{1,j}),
 $$ 
in which the coefficients satisfy $\sum_{j=1}^d |c_{1,j}|^2=1$.
Furthermore, the condition $|\dot\Gamma(s)|=1$ can be satisfied
only if there is a basis in $\R^d$ where $c_{1,1}= ic_{1,2}=
\frac12$ and $c_{1,j}=0$ for $j=3,\dots,d$, in other words, if
$\Gamma$ is a planar circle.
The latter is thus a global maximizer in the class of $C^2$
smooth loops.

It remains to check that the inequality cannot be
saturated for a curve $\Gamma$ that is not $C^2$, so that
the sum (\ref{norm}) diverges.  This would require
 \begin{equation} \label{}
 \frac{\sum_{1\le n\le N} n^2 |c_n|^2
 \left(\frac{\sin\frac{nu}{2}}{n \sin\frac{u}{2}}\right)^2}
{\sum_{1\le n\le N} n^2 |c_n|^2} \,\to\, 1
 \end{equation}
as $N\to\infty$.  This is impossible, however, because
the sum in the numerator is bounded by
$\sec^2\frac{u}{2} \sum_{1\le n\le N} |c_n|^2$ so it
has a finite limit; this concludes the proof. 
\hfill\QED



\setcounter{equation}{0}
\section{An electrostatic isoperimetric problem} \label{elstat}

As a first application of the mean--chord inequalities 
we turn to the electrostatic problem described in the introduction.
Suppose now that $\Gamma$ is a closed $C^2$ curve
in $\R^3$, so that its curvature $\gamma$ and torsion $\tau$ are
continuous functions on $[0,L]$. By $\CC$ we denote a planar
circle; since we work only with distances of pairs of points, it
can be any representative of the equivalence class determined by
Euclidean transformations of $\R^3$. Let us define the quantity
 \begin{eqnarray} \label{potenerg}
 \lefteqn{ \delta(\Gamma):=
 \int_0^L \!\int_0^L \Big[|\Gamma(s) -\Gamma(s')|^{-1} -
 |\CC(s) -\CC(s')|^{-1} \Big] \D s\, \D s'} \nonumber
 \\ &&
 = 2 \int_0^{L/2} \!\!\D u \int_0^L \D s\, \left[
 |\Gamma(s\!+\!u) -\Gamma(s)|^{-1}
 - \frac{\pi}{L}\,\csc \frac{\pi u}{L} \right]\,.
 \end{eqnarray}
Obviously the energy cost of deformation is $q^2\delta(\Gamma)$, where
$q$ is the charge density along the loop, constant by assumption.
The answer to the question stated in the introduction is given by
the following:
\begin{theorem} \label{elstat-iso}
For any closed $C^2$ curve $\Gamma$, 
$\delta(\Gamma)$ is finite and non-negative.  It is zero if and
only if $\Gamma=\CC$, up to Euclidean equivalence.
\end{theorem}
{\sl Proof:} By Theorem~\ref{p=2} and Proposition~\ref{auxil} the
inequality $C_L^{-1}(u)$ is valid for any $u\in(0,\frac12 L]$,
hence the inner integral in the last expression of
(\ref{potenerg}) is non-negative. The trouble, as mentioned in
the introduction, is that taken separately each part leads to a
logarithmically divergent integral due to the behavior of the
integrand as $u\to 0$. It is easy to see, however, that under the
stated regularity requirements on $\Gamma$ one has
 $$ 
 |\Gamma(s\!+\!u) -\Gamma(s)|^{-1} = u^{-1} + \OO(1),
 $$ 
with the error term dependent on $\gamma$ and $\tau$ but uniform
in $s$, hence the singularities cancel and the integral expressing
$\delta(\Gamma)$ converges. \hfill\QED


\setcounter{equation}{0}
\section{An isoperimetric problem for singular interactions} \label{pint}

Let us turn to the singular Schr\"odinger operator (\ref{formal}).
The curve $\Gamma$ is finite, so by \cite{BEKS, BT} we have
$\sigma_\mathrm{ess}(H_{\alpha,\Gamma})= [0,\infty)$ while the
discrete spectrum is nonempty and finite.  In particular,
 $$ 
 \epsilon_1 \equiv \epsilon_1(\alpha,\Gamma):= \inf \sigma
 \left(H_{\alpha,\Gamma}\right)<0\,.
 $$ 
As indicated in the introduction, we ask for which curve
$\Gamma$ the
principal eigenvalue is maximal. The answer is the following.

 \begin{theorem} \label{mainthm}
 Let $\Gamma:\: [0,L]\to \R^2$ have the properties stated in the
 opening of Sec.~\ref{geom}; then
 for any fixed $\alpha>0$ and $L>0$
 $\epsilon_1(\alpha,\Gamma)$
 is  globally uniquely maximized
 by the circle.
 \end{theorem}
{\sl Proof:} The problem reduces to checking the inequality
$C_L^1(u)$ for any $u\in(0,\frac12 L]$ which is achieved by
Theorem~\ref{p=2} and Proposition~\ref{auxil}. The argument was
given in \cite{Ex2}; we recall it briefly here to make the letter
self-contained. One employs a generalized Birman-Schwinger
principle \cite{BEKS} by which there is one-to-one correspondence
between eigenvalues of $H_{\alpha,\Gamma}$ and solutions to the
integral-operator equation
 \begin{equation} \label{bsform}
 \RR_{\alpha,\Gamma}^\kappa \phi= \phi\,, \quad
 \RR_{\alpha,\Gamma}^\kappa(s,s'):= \frac{\alpha}{2\pi}
 K_0\big(\kappa|\Gamma(s) \!-\!\Gamma(s')|\big)\,,
 \end{equation}
on $L^2([0,L])$, where $K_0$ is the Macdonald function; by convention
we write $k=i\kappa$ with $\kappa>0$, corresponding to the
bound-state energy $-\kappa^2$.

To prove the claim it is sufficient to show that 
for any $\kappa>0$
the largest
eigenvalue of the operator $\RR_{\alpha,\Gamma}^\kappa$ given by
(\ref{bsform}) is minimized by a circle.  This
can be combined with the explicit knowledge of the principal
eigenfunction of $\RR_{\alpha,\CC}^\kappa$ which is constant over
the loop, due to the simplicity of the respective eigenvalue
in combination with rotational symmetry. Using the same function
to make a variational estimate in the general case we see that one
has to prove the inequality
 \begin{equation} \label{Greenineq}
 \int_0^L \int_0^L
 K_0\big(\kappa|\Gamma(s) \!-\!\Gamma(s')|\big) \,\D s\D s' \ge
 \int_0^L \int_0^L
 K_0\big(\kappa|\CC(s) \!-\!\CC(s')|\big) \,\D s\D s'
 \end{equation}
for any $\kappa>0$. Using the symmetry of the kernel
and a simple change
of variables we find that (\ref{Greenineq}) is equivalent to
the positivity of the functional
 $$ 
 F_\kappa(\Gamma):= \int_0^{L/2} \D u \int_0^L \D s \left[
 K_0\big(\kappa|\Gamma(s\!+\!u) -\Gamma(s)|\big) -
 K_0\left(\frac{\kappa L}{\pi} \sin \frac{\pi u}{L} \right)
 \right]\,.
 $$ 
By Jensen's inequality,
 $$ 
 \frac 1L \,F_\kappa(\Gamma)\ge \int_0^{L/2} \left[
 K_0\left( \frac{\kappa}{L} \int_0^L |\Gamma(s\!+\!u) -\Gamma(s)| \D s
 \right) - K_0\left(\frac{\kappa L}{\pi} \sin \frac{\pi
 u}{L}\right) \right]\, \D u\,,
 $$ 
where the inequality is strict unless $\int_0^L |\Gamma(s\!+\!u)
-\Gamma(s)| \D s$ is independent of~$s$, because the function
$K_0$ is strictly convex. At the same time, it is decreasing in
$(0,\infty)$, and hence it is sufficient to check that the inequality
 $$ 
 \int_0^L |\Gamma(s\!+\!u) -\Gamma(s)|\, \D s \,\le\,
 \frac{L^2}{\pi} \sin \frac{\pi u}{L}
 $$ 
is valid for all arc lengths $u\in(0,\frac12 L]$ and strict unless
$\Gamma=\CC$; this is nothing else than $C_L^1(u)$.~\hfill\QED

\setcounter{equation}{0}
\section{Mean chords when $p > 2$} \label{chords}

Section 2 provides a complete solution to the problem of maximizing the $L^p$ norm
of the chord of a closed loop of length $2 \pi$ when $p \leq 2$, but 
no information is obtained there for maximizing
$$I(\Gamma,p, u) := \int_0^{2 \pi} {|\Gamma(s\!+\!u) -\Gamma(s)|^p ds} = \|\Gamma(\cdot\!+\!u) -\Gamma(\cdot)\|_p^p$$
with larger values of $p$.  

For any $u, 0 < |u| < \pi$, it is easy to see that $\|\Gamma(s\!+\!u) -\Gamma(s)\|_\infty$
is maximized, with the value $|u|$, by any curve containing a straight interval of length $|u|$
or greater.
Meanwhile, for the circle, $\|\Gamma(s\!+\!u) -\Gamma(s)\|_\infty < u$, 
and therefore, by continuity (\ref{C+p}) is false for sufficiently large values 
of $p$ and any fixed $u, 0 < |u| < \pi$.  The same is true when $u = \pi$, when the 
requirement that $\Gamma \in C^2$ is relaxed.

Little is known, however, about the values of $p>2, u>0$ for which 
the maximizer of $I(\Gamma,p, u)$ remains circular.  
Furthermore, the noncircular 
maximizing curves for large finite $p$ remain unidentified.

Examples show that the isoperimetric result does not extend far beyond the interval $[0, 2]$:

 \begin{example} \label{squished}
 Let $\Gamma$ consist of an interval of length $\pi$ traversed first in one direction and then in the opposite direction.  (This can be regarded as the limiting case of a stadium of perimeter $2 \pi$
 formed by joining two semicircles by parallel straight line segments.)  An elementary 
 calculation shows that for this curve, 
 $\int_0^{2 \pi} {|\Gamma(s\!+\!\pi) -\Gamma(s)|^p\, \D s} = 4 \int_0^{\pi \over 2}{\left(2 x\right)^p \D x}
 = {{2^{2+p}} \over {p+1}} \left({\pi \over 2}\right)^{p+1}$,
 which exceeds the value for the standard circle when $p$ is greater than the largest solution 
 of $p+1 = \left({\pi \over 2}\right)^p$, i.e., numerically for $p \geq 3.15296$.
  \end{example}
  
Example 5.1 provides the smallest value, among simple curves we have checked, of $p$ for which 
(\ref{C+p}) fails.  We note that, paradoxically, examples that are more nearly circular require larger values of 
$p$ for (\ref{C+p}) to fail:
 
 \begin{example} \label{stadium}
 Let $\Gamma_S(a)$ be a standard stadium of length $2 \pi$, for which the straight intervals are of 
 length $\pi a$, supposed small.
A calculation shows that
$$
\left. {{\partial I(\Gamma_S(a),p, \pi)} \over {\partial a}} \right|_{a=0} = 0
$$
and
$$
\left. {{\partial^2 I(\Gamma_S(a),p, \pi)} \over {\partial a}^2} \right|_{a=0}
=  2^{p-1} p \pi \left({\left(\pi^2 - 12\right) + \left(\pi^2 - 8\right)\left({{p \over 2} - 1}\right) }\right),$$
which $> 0$ for $p > {8 \over {\pi^2 - 8}} = 4.27898...$.
\end{example}
 
\begin{example} \label{polygon}
 Let $\Gamma_P(2m)$ be a regular polygon with $2 m$ sides, $m$ supposed large.
A calculation shows that
$$
I(\Gamma_P(2m),p, \pi)
= 2^{1+p} \pi \left|{\sin{u \over 2}}\right|^p \left({1 + p \left({\frac{p-6}{5760}}\right)\left({\pi \over m}\right)^4 + \OO\left(\left({\pi \over m}\right)^6\right) }\right),
$$
which exceeds the value for the circle when $p > 6$ and $m$ is sufficiently large.
\end{example}

The examples lead one to expect that as $p$ increases the maximizing curve may not
bifurcate continuously from the circle, but rather springs from a different geometry
altogether.
Indeed, the circle remains a local 
maximizer of the $p$-mean of the chord, with respect to smooth perturbations,
for all finite $p < \infty$,
in a sense made precise in the following theorem.  For convenience
the curve is placed in $\C$ rather than $\R^2$, entailing a slight shift 
of notation, and a certain choice of orientation is made.
 
 \begin{theorem} \label{LocalStab}
 Let $\Gamma(\gamma, s)$ be a closed curve in $\C$ parametrized by
 arc length $s$, of the form $(1 - \gamma) e^{i s} + \Theta(\gamma,s)$, where $\gamma \geq 0$.  Suppose that $\Theta$ is smooth (say, $C^2$ in $\gamma$ and $s$), and that
 for each $\gamma$, $\Theta(\gamma,s)$ is orthogonal to $e^{i s}$.  Then 
 $\Gamma(0,s)$ is a circle of 
 radius 1, and for any $u$, $0 < u < 2 \pi$,
$$
\left.{{\partial I(\Gamma(\gamma),p, u}) \over {\partial \gamma}} \right|_{\gamma = 0} < 0.
$$
 \end{theorem}
 {\sl Proof:}  The observation (\ref{norm}), which applies to any curve 
 parametrized by arc length, implies 
 that $\Theta(0,s) = 0$, corresponding to the case of a circle.  
Since $\Theta$ is smooth, we may differentiate $I$ 
 under the integral sign, finding
\begin{align*}
 {{\partial I(\Gamma(\gamma),p, u)} \over {\partial \gamma}} &= 
 \int_0^{2 \pi} {{{\partial |\Gamma(\gamma, s\!+\!u) 
-\Gamma(\gamma, s)|^p} \over {\partial \gamma}}\, ds}\\
& = p  \int_0^{2 \pi} |\Gamma(\gamma, s\!+\!u) -\Gamma(\gamma,s)|^{p-2} \\
&\;\;\times
 {\rm Re}\left({\overline{\left({\Gamma(\gamma, s\!+\!u) -\Gamma(\gamma, s)}\right)}
 \left({{{\partial \Gamma(\gamma, s\!+\!u)} \over {\partial \gamma}} -{{\partial \Gamma(\gamma, s)} \over {\partial \gamma}}}\right)}\right) ds.
\end{align*}
When $\gamma = 0$ this simplifies to
\begin{align*}
&p  \left|e^{iu} - 1\right|^{p-2} \int_0^{2 \pi} {\rm Re}
\Biggl( e^{-is}\left({e^{-iu} - 1}\right)\\
&\qquad\times \left({-e^{is}\left({e^{iu} - 1}\right) + 
{{\partial \Theta(0, s\!+\!u)} \over {\partial \gamma}} 
- {{\partial \Theta(0, s)} \over {\partial \gamma}}}\right)\Biggr) ds.
\end{align*}
Since the orthogonality of $e^{i s}$ to $\Theta(\gamma,s)$ implies the 
orthogonality of $e^{i s}$ to $\Theta(\gamma,s+u)$, $\Theta_\gamma(\gamma,s)$, and
$\Theta_\gamma(\gamma,s+u)$,  
 $$
  {{\partial I(\Gamma(0),p, u)} \over {\partial \gamma}} = - p \left|{e^{iu} - 1}\right|^p  \int_0^{2 \pi} {1 ds} = - p 2^{1+p} \pi \left|{\sin{u \over 2}}\right|^p < 0.\eqno\QED
 $$


\subsection*{Acknowledgments}

P.E. is grateful for the hospitality at Georgia Tech where a part
of the work was done. The research was supported by ASCR and its
Grant Agency within the projects IRP AV0Z10480505 and A100480501;
and by US NSF grants DMS-0204059 and DMS-0300349.
%


\begin{thebibliography}{99}

 \bibitem{AGHH2}
 S.~Albeverio, F.~Gesztesy, R.~H\o egh-Krohn, H.~Holden: {\em
 Solvable Models in Quantum Mechanics}, 2nd printing with an
 appendix by P.~Exner, AMS Chelsea Publ., Providence, R.I., 2005.
 \vspace{-1.8ex}
 \bibitem{BEKS}
 J.F.~Brasche, P.~Exner, Yu.A.~Kuperin, P.~\v{S}eba: Schr\"odinger
 operators with singular interactions, {\em J. Math. Anal. Appl.}
 {\bf 184} (1994), 112-139. \vspace{-1.8ex}
 \bibitem{BT}
 J.F.~Brasche, A.~Teta: Spectral analysis and scattering theory for
 Schr\"odinger operators with an interaction supported by a regular
 curve, in {\em Ideas and Methods in Quantum and Statistical
 Physics}, Cambridge Univ. Press 1992; pp.~197-211. \vspace{-1.8ex}
 \bibitem{Ex1}
 P.~Exner: An isoperimetric problem for point interactions, {\em J.
 Phys. A: Math. Gen.} {\bf A38} (2005), 4795-4802.
 \vspace{-1.8ex}
 \bibitem{Ex2}
 P.~Exner: An isoperimetric problem for leaky loops and
related mean-chord inequalities, {\em J. Math. Phys.} {\bf 46}
(2005), 062105
 \vspace{-1.8ex}
 \bibitem{EHL}
 P.~Exner, E.M.~Harrell, M.~Loss: Optimal eigenvalues for some
 Laplacians and Schr\"odinger operators depending on curvature, in
 {\em Operator Theory : Advances and Applications}, vol.~108;
 Birk\-h\"auser Verlag, Basel 1998; pp.~47--53.
 \vspace{-1.8ex}
 \bibitem{EI}
 P.~Exner, T.~Ichinose: Geometrically induced spectrum in curved
 leaky wires, {\em J. Phys. A: Math. Gen.} {\bf 34}, 1439--1450 (2001).
 \vspace{-1.8ex}
 \bibitem{KF}
 A.N.~Kolmogorov, S.V.~Fomin: \emph{Elements of Function Theory
 and Functional Analysis}, 3rd edition, Nauka, Moscow 1972
 \vspace{-1.8ex}

\end{thebibliography}
\end{document}